\documentclass[12pt]{article}
\usepackage{float} 
\usepackage{amsmath}
\usepackage{slashed}
\usepackage{amsfonts}   
\usepackage{amssymb}

\begin{document}
\date{}
\title{{\bf{\Large Penrose limit for holographic duals of $ J\bar{T} $ deformations}}}
\author{
 {\bf {\normalsize Dibakar Roychowdhury}$
$\thanks{E-mail:  dibakarphys@gmail.com, dibakar.roychowdhury@ph.iitr.ac.in}}\\
 {\normalsize  Department of Physics, Indian Institute of Technology Roorkee,}\\
  {\normalsize Roorkee 247667, Uttarakhand, India}
\\[0.3cm]
}

\maketitle
\begin{abstract}
We explore bosonic string sigma models on warped $ BTZ\times S^3 $ both in the plane wave as well as beyond plane wave limit. Using the light cone gauge, we obtain the corresponding Hamiltonian and therefore the spectrum associated with the pp wave strings. Our analysis reveals a constant shift in the spectrum arising as a result of TsT transformations along the isometries of the target space manifold. Imposing the energy positivity constraint on the CFT$ _2 $ spectrum, we estimate an upper bound on the shift. We also estimate corrections as we go beyond the plane wave limit. Finally, we perform calculations using conformal gauge which reveals identical spectrum for the pp wave strings and thereby shows the equivalence between the two approaches.
\end{abstract}
\section{Overview and Motivation}
For many years, explorations of stringy dynamics on AdS$ _3 $ \cite{Giveon:1998ns}-\cite{Argurio:2000tb} and beyond \cite{Azeyanagi:2012zd} has sharpened our current understanding of $ AdS_3/CFT_2 $ duality in a significant way. Recently, the ongoing developments along this path has received a significant boost in the light of irrelevant deformations in 2D CFTs \cite{Zamolodchikov:2004ce}-\cite{Guica:2017lia}. These deformations are special in the sense that the spectrum of the deformed theory could be obtained in terms of the original (undeformed) CFT data. For the purpose of this paper, we will be concerned about the $ J\bar{T} $ deformations \cite{Guica:2019vnb}-\cite{Guica:2017lia} and in particular its stringy counterpart at zero temperature.

On the stringy side of the correspondence, these deformations are realized as marginal current-current deformations of the string world-sheet on  $ AdS_3\times S^3 \times \mathcal{M}^4 $ \cite{Hashimoto:2019wct}-\cite{Bzowski:2018pcy}. The recent analysis of \cite{Apolo:2019yfj}-\cite{Apolo:2019zai} reveals that marginal deformations on the string world-sheet could be realized as TsT transformations along the target space coordinates in the presence of NS-NS fluxes. It turns out that in case of $ J\bar{T} $ deformations, one of the isometry directions belongs to the $ AdS_3 $ factor while the other $ U(1) $ direction comes from $ S^3 $ \cite{Apolo:2019yfj}.

Given the above correspondence, the purpose of the present paper is to explore the Penrose limit for bosonic sigma models over warped $ BTZ\times S^3 $ in the presence of NS-NS fluxes. In particular, we present a systematic way of taking the pp wave limit \cite{Berenstein:2002jq}-\cite{Gava:2002xb} for warped \emph{massless} $ BTZ \times S^3 $ spacetime which unveils three flat directions associated with the target space manifold. Two of these directions give rise to massive stringy modes upon quantization while the other remains as the massless excitation.

The quantization of the corresponding light cone Hamiltonian predicts a shift
\begin{eqnarray}
\Delta_S \sim \frac{\tilde{\gamma}\mathfrak{p}^+}{\omega}
\end{eqnarray}
in the spectrum of the 2D CFT whose source (in the bulk theory) lies in the TsT transformations acting along the isometries of the target space manifold. Here, $ \tilde{\gamma} $ is the reminiscent of the TsT deformation parameter in the pp wave limit of the bosonic sigma model. On the other hand, $ \mathfrak{p}^{+} $ is the momentum associated with the string along light cone direction $ \mathfrak{x}^+ $ and $ \omega $ is the mass parameter of the model which we later identify as proportional to the rest mass of the massive stringy modes associated with the sigma model. Imposing the energy positivity condition ($ \Delta >0 $) on the spectrum, we further estimate an upper bound on the TsT transformation parameter ($ \tilde{\gamma} $) and thereby on the energy shift $ \Delta_S $.

As a further continuation of our analysis, we go beyond the conventional pp wave limit \cite{Callan:2003xr}-\cite{Callan:2004uv} and estimate $ \ell^{-2} $ corrections to the pp wave Hamiltonian (and thereby the spectrum) where $ \ell $ is the length scale associated to $ AdS_3 $. We also construct the pp wave Hamiltonian using conformal gauge and show the equivalence between the Hamiltonian densities in two approaches. Finally, we draw conclusion in Section 3.
\section{Analysis and results}
\subsection{The background solution}
Let us begin by discussing the \emph{massless} BTZ solution prior to the application of any TsT transformation. The full bosonic sector of the solution is of the form $ AdS_3 \times S^3 \times \mathcal{M}_4 $ supported by NS-NS flux. However, for the purpose of our current discussion, we retain ourselves only to the $ AdS_3 $ part of the full bosonic solution.

The metric corresponding to the BTZ black holes is represented by the line element
\begin{eqnarray}
\label{btz}
ds^2 = \frac{\ell^2 dr^2}{(r^2-r^2_+)}+\frac{r^2}{\ell^2}d\varphi^2 -\frac{(r^2-r^2_+)}{\ell^2}dt^2,
\end{eqnarray}
where the corresponding mass as well as the temperature of the solution (\ref{btz}) is expressed in terms of the horizon radius namely, $ M=\frac{r^2_+}{8G_3 \ell^2} $ and  $T_H=\frac{r_+}{2\pi \ell^2}$.  

For the purpose of our analysis, we will look for a particular limit of (\ref{btz}) where we set $ r_+=0 $. This limit is known as the massless ($ M=0 $) limit of BTZ solution (\ref{btz}) which represents a zero temperature ($ T_H =0 $) background of the form
\begin{eqnarray}
\label{massless}
ds^2 \sim \frac{dr^2}{r^2}+r^2 (d\varphi^2 -dt^2)+d\Omega^2_3 (\chi , \psi , \theta),
\end{eqnarray}
where, $ d\Omega^2_3 $ represents the line element on $ S^3 $.

The singularity near $ r \sim 0 $ of the massless solution (\ref{massless}) is an artefact of the existence of a \emph{conical} defect in corresponding $ 3d $ Euclidean version of the solution. The conical singularity arises due to the presence of the contractible cycle along the time direction. As we shall see, this feature persists even after the application of TsT to the above background solution (\ref{massless}). In other words, the TsT transformation does not seem to remove the original conical defect present with the massless BTZ solution (\ref{massless}).

The TsT transformation, when acted upon on (\ref{massless}) gives rise to what is known as \emph{warped} massless BTZ spacetime. Below we briefly summarise the TsT application procedure which in fact reveals that during its application, the TsT operation does not seem to touch the $ r $ direction and thereby the singularity associated to it. 

The TsT transformation consists of a set of T duality transformations accompanied with a shift \cite{Apolo:2019yfj}-\cite{Apolo:2019zai}.  The first T duality is acted upon on the isometric direction $ \chi \rightarrow \chi' $ which is followed by a corresponding shift, $ v' \rightarrow t^-  +\frac{2\gamma}{k}\tilde{\psi}$ and $ \chi' \rightarrow \tilde{\psi} $ where, $ v=\varphi -t $ is null coordinate associated to (\ref{massless}). Finally, we apply a second T duality on $ \tilde{\psi} $ which together with the additional change in coordinates, $ t^- = v+\frac{\gamma}{2}\chi $ and $ \tilde{\psi} =\chi $ yields a background solution (known as warped massless BTZ) of the form  \cite{Apolo:2019yfj}
\begin{eqnarray}
\label{e1}
ds^2&=&\ell^2 \left\lbrace \frac{dr^2}{4r^2}+r(d\varphi^{2}-dt^2)+\gamma r (d \varphi +d t)(d \chi + \cos\theta d\psi)+d\Omega_{3}^2\right\rbrace ,\\
d\Omega_{3}^2&=&\frac{1}{4}\left((d \chi + \cos\theta d\psi)^2 +d\theta^2 +\sin^2\theta d\psi^2\right), \\
\mathfrak{B}_{2}&=&\ell^2 \left\lbrace \frac{1}{4}\left(\cos\theta d\psi\wedge d\chi +4r d\varphi \wedge dt \right) -\frac{\gamma r}{2}(d \varphi +dt)\wedge (d \chi + \cos\theta d\psi)\right\rbrace ,
\label{e3}
\end{eqnarray}
where, $ \gamma $ stands for the TsT deformation parameter and $ \ell = \sqrt{k} \ell_s $ defines the radius of curvature of $ AdS_3 $. Here, $ \ell_s=\sqrt{\alpha'} $ defines the string length and $ k $ defines the level of the world-sheet current algebra \cite{Giveon:1998ns}-\cite{Kutasov:1999xu}.

Interestingly enough, one can show that the TsT transformation acted upon on the target space coordinates of the massless BTZ (\ref{massless}) could be recast as \emph{marginal} current-current deformation on the string world-sheet propagating over massless $ BTZ \times S^3 $  \cite{Apolo:2019yfj}. Marginal deformation of the above type corresponds to single trace (irrelevant) $ J\bar{T} $ deformation of the dual CFT living in $ 2d $ \cite{Apolo:2018qpq}-\cite{Chakraborty:2018vja}. Therefore, connecting all these pieces together, one can conclude that warped massless BTZ solution of the above type (\ref{e1})-(\ref{e3}) are dual to $ J\bar{T} $ deformed $ 2d $ CFT at zero temperature. The finite temperature version of these CFTs are realised in terms of warped BTZ black strings \cite{Argurio:2000tb},\cite{Apolo:2019yfj}.
\subsection{pp wave dynamics}
\subsubsection{Towards Penrose limit}
To proceed further, we first introduce the following change in coordinates,
\begin{eqnarray}
\chi &=& \Psi + \Phi ~;~\psi = \Psi - \Phi ~;~\theta = 2\Theta ~;~ r= \cosh^2\rho
\label{e4}
\end{eqnarray} 
which finally yields the background space-time metric and the NS-NS flux,
\begin{eqnarray}
\frac{ds^2}{\ell^2}=\tanh^2\rho d\rho^2 +\cosh^2\rho(d\varphi^{2}-dt^2)+d \Theta^2 + \sin^2 \Theta d\Phi^2  \nonumber\\
+\cos^2\Theta d \Psi^2 + 2\gamma \cosh^2\rho (d \varphi +  dt) (\cos^2 \Theta d \Psi + \sin^2\Theta d\Phi)
\label{e5}
\end{eqnarray}
\begin{eqnarray}
\frac{\mathcal{B}_{2}}{\ell^2}= \frac{1}{2}\cos 2\Theta d\Psi \wedge d\Phi +\cosh^2\rho d\varphi \wedge dt \nonumber\\
- \gamma \cosh^2\rho (d\varphi + dt) \wedge (\cos^2 \Theta d \Psi +\sin^2\Theta d\Phi).
\label{e6}
\end{eqnarray}

It is important to notice that, unlike previous examples \cite{Gomis:2002qi}-\cite{Gava:2002xb}, the pp wave limit should be considered by zooming into the geometry around null geodesics at a finite radial distance $ r=r_c +\mathcal{O}(\ell^{-2})$ from the centre ($ r=0 $) of the warped $ BTZ $ spacetime. This is because any expansion around $ r \sim 0 $ is ill defined due to the presence of the singularity. 

As elaborated earlier, this is a characteristic feature of all \emph{massless} BTZ solutions and therefore also applies to (\ref{e1}) which might be regarded as the zero temperature limit of its finite temperature counterpart known as the warped BTZ black strings \cite{Argurio:2000tb},\cite{Apolo:2019yfj}. In other words, the above solution (\ref{e1}) might be regarded as the ground state of warped BTZ black strings with $ M=0 $ and $ T_H=0 $. This is similar in spirit to the example of massless BTZ (\ref{massless}) which is the ground state of the BTZ black hole as introduced in (\ref{btz}).

After taking into account all these facts, a careful analysis reveals that at the pp wave limit, the target space geometry effectively looks like a 5D manifold with three flat directions attached. However, non trivial radial dependencies starts showing up as we go beyond the traditional pp wave limit \cite{Callan:2003xr}-\cite{Callan:2004uv} and therefore the resulting geometry modifies significantly.

 We introduce the pp wave limit of (\ref{e5})-(\ref{e6}) by redefining coordinates as,
\begin{eqnarray}
t =\omega \mathfrak{x}^{+}~;~\varphi = \frac{\eta}{\ell}~;~\Psi = \omega \mathfrak{x}^{+}-\frac{\mathfrak{x}^{-}}{\omega \ell^2}~;~\rho =\frac{\varrho}{\ell}~;~\Theta = \frac{\beta}{\ell}~;~\gamma = \frac{\tilde{\gamma}}{\omega^2 \ell^2}
\label{e7}
\end{eqnarray}
and thereby taking $ \ell \rightarrow \infty $ keeping $ \ell_s= $ fixed. The above corresponds to the limit $ k \rightarrow \infty $ in which the world-sheet sigma model on $ AdS_3$ becomes weakly coupled. This is the limit which sets, $ \rho \rightarrow 0 $ and thereby fixes the radial coordinate as, $ r=1+\mathcal{O}(\ell^{-2}) $. 

Using (\ref{e7}), the pp wave limit of the metric (\ref{e5}) turns out to be,
\begin{eqnarray}
ds^2_{pp}&=&- 2 d\mathfrak{x}^{+}d\mathfrak{x}^{-}-(\beta ^2 \omega ^2  - 2 \tilde{\gamma} )(d\mathfrak{x}^{+})^2+d\eta^2 +d\beta^2+\beta ^2 d\Phi^2 \nonumber\\
&=&- 2 d\mathfrak{x}^{+}d\mathfrak{x}^{-}-( \omega ^2 \mathcal{X}^2_a  - 2 \tilde{\gamma} )(d\mathfrak{x}^{+})^2+d\mathcal{Z}^2_i ~;~a,b=1,2
\label{e8}
\end{eqnarray}
where, we redefine our coordinates as,
\begin{eqnarray}
\mathcal{X}_1 = \beta \cos\Phi ~;~ \mathcal{X}_2 = \beta \sin\Phi ~;~\mathcal{Z}_i = \lbrace\eta , \mathcal{X}_a= \mathcal{X}_1 , \mathcal{X}_2\rbrace.
\end{eqnarray}

The NS-NS flux, on the other hand, needs to be normalized in the pp wave limit,
\begin{eqnarray}
\mathcal{B}_{2pp}\sim \frac{\mathcal{B}_2}{\omega^2 \ell^2}&=&\frac{1}{2 \omega}d\mathfrak{x}^{+}\wedge d\Phi.
\label{e10}
\end{eqnarray}
\subsubsection{pp wave Hamiltonian}
Given the pp wave background (\ref{e8}), our first task is to identify the conserved charges associated with the sigma model in the light cone coordinates ($ \mathfrak{x}^{\pm}$). A straightforward computation reveals string theory charges as,
\begin{eqnarray}
\mathfrak{p}^{-}&=&i \partial_{\mathfrak{x}^+}= 2\omega (\Delta -J)\sim  E_S - J\\
\mathfrak{p}^{+}&=&i \partial_{\mathfrak{x}^-}=\frac{1}{\omega \ell^2}(\Delta +J)\sim E_S + J
\label{eqn13}
\end{eqnarray}
where, $ E_S(\sim i \partial_t) $ is the classical energy of the string that stands for the conformal dimension ($ \Delta $) of the operator in the dual CFT$ _2 $. On the other hand, $ J (\sim -i \partial_{\Psi})$ is the $ R $- charge in terms of CFT$ _2 $ d.o.f. such that the operators satisfy the BPS bound $ \Delta \geq J $.

The sigma model describing (semi)classical strings propagating over the pp wave geometry (\ref{e8}) is given by,
\begin{eqnarray}
\mathcal{S}_{pp}=-\frac{1}{4 \pi \alpha'}\int d^2 \sigma \mathcal{L}_{pp} 
\end{eqnarray}
where the bosonic part of the Lagrangian density is given by\footnote{At the level of the sigma model, the NS-NS contribution appears to be a total derivative term which therefore can be ignored in the strict pp wave limit.},
\begin{eqnarray}
\mathcal{L}_{pp}=\mathfrak{h}^{\alpha \beta}\partial_{\alpha}X^{M}\partial_{\beta}X^N \mathcal{G}_{MN}-\varepsilon^{\alpha \beta}\partial_{\alpha}X^{M}\partial_{\beta}X^N \mathcal{B}_{MN}.
\label{e14}
\end{eqnarray}
Here, $ \mathfrak{h}^{\alpha \beta}=\sqrt{-\det\zeta_{\alpha \beta}}~\zeta^{\alpha \beta} $ where $ \zeta_{\alpha \beta} $ is the 2D world-sheet metric. On the other hand, $ \varepsilon^{\alpha \beta} $ is the 2D Levi-Civita symbol with the convention $ \varepsilon^{\tau \sigma}=-\varepsilon^{\sigma \tau} =1$.

To proceed further, we (gauge) fix the reparametrization invariance of the world-sheet theory by adopting the so called light cone gauge \cite{Asano:2015qwa}-\cite{Parnachev:2002kk},
\begin{eqnarray}
\mathfrak{x}^{+}=\tau ~;~ \partial_{\sigma}\zeta_{\sigma \sigma}=0~;~\sqrt{-\det\zeta_{\alpha \beta}}=1
\end{eqnarray}
which upon substitution into (\ref{e14}) yields,
\begin{eqnarray}
\mathcal{L}_{pp}=\zeta_{\sigma \sigma}(\tau)\left(2 \partial_{\tau}\mathfrak{x}^{-}-(\partial_{\tau}\mathcal{Z}_i)^2 +\omega^{2}\mathcal{X}^2_a -2 \tilde{\gamma} \right) -2\zeta_{\tau \sigma}(\tau ,\sigma)(\partial_{\sigma}\mathfrak{x}^{-}-\partial_{\tau}\mathcal{Z}_i \partial_{\sigma}\mathcal{Z}_i)\nonumber\\
+\zeta_{\sigma \sigma}^{-1}(\tau)(1-\zeta_{\tau\sigma}^2(\tau , \sigma))(\partial_{\sigma}\mathcal{Z}_i)^2.
\label{e16}
\end{eqnarray}

The canonically conjugate momenta ($ \mathfrak{p}_{M} $) are given below\footnote{Following (\ref{eqn13}), one can show that $ \mathfrak{p}_{-}=g_{+-}\mathfrak{p}^{+}=- \mathfrak{p}^{+}$ is a constant.},
\begin{eqnarray}
\label{e17}
\mathfrak{p}_{-}&=&2 \zeta_{\sigma \sigma}, \\
\mathfrak{p}_{i}&=&-2 \zeta_{\sigma \sigma} \partial_{\tau}\mathcal{Z}_i + 2 \zeta_{\tau \sigma}\partial_{\sigma}\mathcal{Z}_i.
\label{e18}
\end{eqnarray}

On the other hand, the Virasoro constraints follow directly from the stress energy tensor associated with the 2D world-sheet theory,
\begin{eqnarray}
T_{\alpha \beta}=\partial_{\alpha}X^M \partial_{\beta}X^N \mathcal{G}_{MN}-\frac{1}{2}\zeta_{\alpha \beta}\zeta^{\mu \nu}\partial_{\mu}X^M \partial_{\nu}X^N \mathcal{G}_{MN}=0.
\end{eqnarray}

Setting, $ \alpha = \tau $ and $ \beta= \tau $ we find
\begin{eqnarray}
\partial_{\tau}\mathfrak{x}^{-}+\frac{1}{2}(\omega^2 \mathcal{X}^2_a -2 \tilde{\gamma})-\frac{1}{2 \mathfrak{p}^2_-}(2\zeta_{\tau \sigma}\partial_{\sigma}\mathcal{Z}_i-\mathfrak{p}_i)^2-\frac{2}{\mathfrak{p}^2_-}\frac{(1- \zeta^2_{\tau \sigma})^2}{(1+\zeta^2_{\tau \sigma})}(\partial_{\sigma}\mathcal{Z}_i)^2\nonumber\\
+\frac{2}{\mathfrak{p}_-}\frac{\zeta_{\tau \sigma}(1-\zeta^2_{\tau \sigma})}{(1+ \zeta^2_{\tau \sigma})}\left(\partial_{\sigma}\mathfrak{x}^{-}+\frac{\partial_{\sigma}\mathcal{Z}_i}{\mathfrak{p}_-}(\mathfrak{p}_i - 2\zeta_{\tau \sigma}\partial_{\sigma}\mathcal{Z}_i) \right) =0.
\label{e20}
\end{eqnarray}

On the other hand, setting $ \alpha = \tau $ and $ \beta= \sigma $ we find
\begin{eqnarray}
\partial_{\tau}\mathfrak{x}^{-}+\frac{1}{2}(\omega^2 \mathcal{X}^2_a -2 \tilde{\gamma})-\frac{1}{2 \mathfrak{p}_-}(2\zeta_{\tau \sigma}\partial_{\sigma}\mathcal{Z}_i-\mathfrak{p}_i)\partial_{\sigma}\mathcal{Z}_i+\frac{2}{\mathfrak{p}_{-}}\frac{(1-\zeta^2_{\tau \sigma})\partial_{\sigma}\mathfrak{x}^{-}}{\zeta_{\tau \sigma}}\nonumber\\
-\frac{2}{\mathfrak{p}^2_-}\frac{(1-\zeta^2_{\tau \sigma})}{\zeta_{\tau \sigma}}(2\zeta_{\tau \sigma}\partial_{\sigma}\mathcal{Z}_i-\mathfrak{p}_i)\partial_{\sigma}\mathcal{Z}_i+\frac{2}{\mathfrak{p}^2_{-}}(1-\zeta^2_{\tau \sigma})(\partial_{\sigma}\mathcal{Z}_i)^2=0.
\label{e21}
\end{eqnarray}

Solving (\ref{e20}) and (\ref{e21}) we find,
\begin{eqnarray}
\partial_{\tau}\mathfrak{x}^{-}=-\frac{1}{2}(\omega^2 \mathcal{X}^2_a -2 \tilde{\gamma})-\frac{\zeta^2_{\tau \sigma}}{2 \mathfrak{p}_-}(2\zeta_{\tau \sigma}\partial_{\sigma}\mathcal{Z}_i-\mathfrak{p}_i)\partial_{\sigma}\mathcal{Z}_i \nonumber\\
+\mathfrak{h}(\tau , \sigma)(2\zeta_{\tau \sigma}\partial_{\sigma}\mathcal{Z}_i - \mathfrak{p}_i)^2+\mathfrak{g}(\tau , \sigma)(\partial_{\sigma}\mathcal{Z}_i)^2
\end{eqnarray}
and,
\begin{eqnarray}
\partial_{\sigma}\mathfrak{x}^{-}=\left( \frac{1}{\mathfrak{p}_{-}}+\frac{\mathfrak{f}(\tau , \sigma)}{4}\right) (2\zeta_{\tau \sigma}\partial_{\sigma}\mathcal{Z}_i - \mathfrak{p}_i)\partial_{\sigma}\mathcal{Z}_i - \frac{2}{\mathfrak{p}_-}\zeta_{\tau \sigma}(\partial_{\sigma}\mathcal{Z}_i)^2\nonumber\\
-\frac{\mathfrak{f}(\tau , \sigma)}{4 \mathfrak{p}_-}(2\zeta_{\tau \sigma}\partial_{\sigma}\mathcal{Z}_i - \mathfrak{p}_i)^2
\label{e23}
\end{eqnarray}
where we introduce new variables as,
\begin{eqnarray}
\mathfrak{f}(\tau , \sigma)&=& \frac{\zeta_{\tau \sigma}(1+\zeta^2_{\tau \sigma})}{1-\zeta^2_{\tau \sigma}}~;~\mathfrak{h}(\tau , \sigma)=\frac{(1+ \zeta^2_{\tau \sigma})}{2 \mathfrak{p}^2_-}\\
\mathfrak{g}(\tau , \sigma)&=&\frac{2}{\mathfrak{p}^2_{-}}(1-\zeta^2_{\tau \sigma}).
\end{eqnarray}

Finally, collecting all the pieces together the pp wave Hamiltonian density in the light cone gauge turns out to be,
\begin{eqnarray}
\mathcal{H}^{(lc)}_{pp}\equiv -\mathfrak{p}_+=-\frac{1}{4 \mathfrak{p}_-}(2\zeta_{\tau \sigma}\partial_{\sigma}\mathcal{Z}_i - \mathfrak{p}_i)^2 - \frac{\mathfrak{p}_-}{4}(\omega^{2}\mathcal{X}^2_a -2 \tilde{\gamma})\nonumber\\
+\zeta_{\tau \sigma}\partial_{\sigma}\mathfrak{x}^{-}
-\frac{1}{\mathfrak{p}_-}(1-\zeta^2_{\tau \sigma})(\partial_{\sigma}\mathcal{Z}_i)^2
\end{eqnarray}
where $ \partial_{\sigma}\mathfrak{x}^{-} $ may be replaced using (\ref{e23}).

The equation of motion corresponding to $ \mathfrak{x}^{-} $ together with the input from (\ref{e17}) implies that, $ \partial_{\sigma}\zeta_{\tau \sigma}=0 $. Using the residual gauge (Weyl) freedom on the world-sheet \cite{Parnachev:2002kk} we can further fix $ \zeta_{\tau \sigma}=0 $ which thereby yields the Hamiltonian of the pp wave string as, 
\begin{eqnarray}
\mathfrak{H}_{pp}=\frac{1}{2\pi}\int_{0}^{2\pi} d\sigma \left(\frac{\mathfrak{p}^2_a}{4\mathfrak{p_-}}+\frac{\mathfrak{p}_-}{4}(\omega^{2}\mathcal{X}^2_a -2 \tilde{\gamma})+\frac{1}{\mathfrak{p}_-}(\partial_{\sigma}\mathcal{X}_a)^2+ \frac{\mathfrak{p}^2_{\eta}}{4\mathfrak{p_-}}+\frac{1}{\mathfrak{p}_-}(\partial_{\sigma}\eta)^2\right) 
\end{eqnarray}
which shows that $ \mathcal{X}_a $s are the only \emph{massive} excitations with rest mass $ \omega\sim \mathfrak{m} $. 
\subsubsection{The spectrum}
In order to compute the spectrum of the theory, we propose the following mode expansion \cite{Parnachev:2002kk} for the world-sheet fields,
\begin{eqnarray}
\label{e28}
\mathcal{X}_{a}(\tau , \sigma)&=& \sum_{n}\frac{1}{\sqrt{\mathfrak{w}_{n}}}\left(\mathfrak{c}_a^n  e^{-\frac{i}{\mathfrak{p}_-}(\mathfrak{w}_{n}\tau +\mathfrak{K}_n \sigma)} + c.c. \right)\\
\eta (\tau , \sigma)&=&  \sum_{n}\frac{1}{\sqrt{\nu_{n}}}\left(\mathfrak{b}^n_{(\eta)}  e^{-\frac{i}{\mathfrak{p}_-}(\nu_{n}\tau +\kappa_n \sigma)} + c.c. \right).
\label{e29}
\end{eqnarray}

Using (\ref{e28}) and (\ref{e29}) and following the normal ordering the spectrum may finally be expressed as,
\begin{eqnarray}
\Delta =J+ \Delta_{S}+\frac{1}{ 2 \omega \mathfrak{p}_-}\sum_n \nu_{n}\mathcal{N}_{\eta}(\kappa_n)+\frac{2}{\mathfrak{p}^2_-}\sum_{n,a}\mathcal{N}_{a}(\mathfrak{K}_{n}) \nonumber\\
+\frac{1}{\mathfrak{m}^2\mathfrak{p}^2_-}\sum_{n ,a}\mathfrak{K}^2_n \mathcal{N}_{a}(\mathfrak{K}_{n}) +\mathcal{O}(1/\mathfrak{p}^3_-)
\label{e31}
\end{eqnarray}
where we identify the number operators corresponding to massless as well as massive modes respectively as,
\begin{eqnarray}
\mathcal{N}_{\eta}(\kappa_{n})=\mathfrak{b}_{(\eta)}^{n\dagger}\mathfrak{b}^n_{(\eta)}
\end{eqnarray}
and,
\begin{eqnarray}
\mathcal{N}_{a}(\mathfrak{K}_{n})=\mathfrak{c}_{a}^{n\dagger}\mathfrak{c}^n_{a}.
\end{eqnarray}

On the other hand, the corresponding dispersion relations for these modes could be respectively expressed as,
\begin{eqnarray}
\nu^2_n = \kappa^2_n
\end{eqnarray}
and
\begin{eqnarray}
\mathfrak{w}_{n}^2=\mathfrak{K}^2_n +\mathfrak{m}^2~;~\mathfrak{m}^2 = \omega^2 \mathfrak{p}^2_-=4\omega^2
\end{eqnarray}
subjected to the fact that we set, $ \mathfrak{p}_- =2 $ in order for $ \eta $ to be \emph{massless} mode. This naturally fixes $ \zeta_{\sigma \sigma} =1$ by virtue of (\ref{e17}). Finally, we notice that the effect of TsT on target space coordinates (or equivalently the marginal deformation on the 2D world-sheet theory) appears as a constant shift,
\begin{eqnarray}
\label{eqn36}
\Delta_{S}= \frac{\tilde{\gamma}}{4 \omega}\mathfrak{p}^{+}
\end{eqnarray} 
in the spectrum where $ \mathfrak{p}^{+} $ is the light cone momentum of the string.

The above result (\ref{eqn36}), as we elaborate below, comprises of all the terms in the form of $ 1/J $ expansion in the large $ J( \rightarrow \infty) $ limit. The leading term in the expansion corresponds to the point particle limit of the string where all the quantum fluctuations are suppressed due to large $ J $ effects. However, quantum fluctuations start contributing in the form of massless as well as massive modes as soon as we start estimating the $ 1/J $ corrections. 
\subsubsection{A bound on $ \Delta_S $}
From (\ref{e31}), one can further extract an upper bound on the TsT deformation parameter ($ \tilde{\gamma} $) and hence in general on the shift ($ \Delta_S $) in the pp wave spectrum. A careful analysis reveals that one could think of (\ref{e31}) as an expansion in the \emph{effective} (large) $R$- charge $ \tilde{J}(=J -\frac{\tilde{\gamma}}{2})$ where we drop all the sub leading terms in $ \frac{\omega}{\tilde{J} \mathfrak{p}^+}$ being very small\footnote{The above formula (\ref{eqn37}) can be recast as the large $ J (\rightarrow \infty) $ expansion corresponding to a plane wave background \cite{Berenstein:2002jq}. This stems from the fact that, $ \mathfrak{p}^{+} \sim E_S + J $ (\ref{eqn13}). In particular, the above form (\ref{eqn37}) is quite pertinent to the plane wave strings those are dual to $ J\bar{T} $ deformed CFTs. It simply says that the entity within the square root has to be positive definite in order for the theory to make sense. As we see, this provides an upper bound on the TsT parameter ($ \gamma $) beyond which the energy becomes complex. Similar observations were made by authors in \cite{Apolo:2019yfj} without actually going into the plane wave limit.},
\begin{eqnarray}
\label{eqn37}
\Delta &\approx &\tilde{J}\sqrt{1-\frac{2 \mathfrak{n}_{\nu} \omega}{\tilde{J}\mathfrak{p^+}}~\jmath (\mathfrak{p}^+)}\\
\mathfrak{n}_{\nu} &=&\frac{1}{ 2 \omega }\sum_n \nu_{n}\mathcal{N}_{\eta}(\kappa_n).
\end{eqnarray}

The function $ \jmath (\mathfrak{p}^+) $, on the other hand, could be formally expressed as,
\begin{eqnarray}
 \jmath (\mathfrak{p}^+) = 1-\frac{\mathfrak{n}_{\mathfrak{K}}}{\mathfrak{n}_{\nu}}\frac{\omega}{\mathfrak{p}^+}+\cdots
\end{eqnarray}
where we identify,
\begin{eqnarray}
\mathfrak{n}_{\mathfrak{K}} = 2\sum_{n,a}\mathcal{N}_{a}(\mathfrak{K}_{n}) 
+\frac{1}{\mathfrak{m}^2}\sum_{n ,a}\mathfrak{K}^2_n \mathcal{N}_{a}(\mathfrak{K}_{n}). 
\end{eqnarray}

Now, demanding the energy positivity ($ \Delta> 0 $) condition on the spectrum, we find the following upper bound on the TsT deformation parameter,
\begin{eqnarray}
\label{eqn41}
\tilde{\gamma}\leq \tilde{\gamma}_{c}= 2J -\frac{4\mathfrak{n}_{\nu}\omega}{\mathfrak{p}^+}+\mathcal{O}(1/(\mathfrak{p}^+)^2).
\end{eqnarray}

Substituting (\ref{eqn41}) into (\ref{eqn36}) we find an upper bound on the corresponding energy shift,
\begin{eqnarray}
\label{eqn42}
\Delta_S \leq \Delta^{(c)}_S= \frac{\mathfrak{p}^+ J}{2 \omega}-\mathfrak{n}_{\nu}+\mathcal{O}(1/\mathfrak{p}^+)
\end{eqnarray}
where, $ \mathfrak{n}_{\nu} $ is the total occupation number corresponding to zero modes in the spectrum.

The above upper bound (\ref{eqn41}) is an artifact of the existence of a maximal (right moving) energy in the $ J\bar{T} $ deformed CFT beyond which the spectrum becomes complex \cite{Apolo:2019yfj}. In the plane wave limit, the stringy counterpart of this bound is precisely realised in the form of (\ref{eqn42}). On the other hand, the lower bound of the deformation parameter (\ref{eqn41}) may be set equal to zero which maps the right moving energy of the $ J\bar{T} $ deformed CFT to the right moving energy of the undeformed CFT. 

Notice that, in the above expression (\ref{eqn41}) we have ignored corrections beyond $ 1/\mathfrak{p}^+$ which is true only in the limit of the large angular momentum $ J \rightarrow \infty $. In other words, all the sub-leading corrections are smaller compared to the leading term in the expansion (\ref{eqn41}). This is the large $ J $ expansion which is precisely justified in the plane wave limit \cite{Berenstein:2002jq}. 

However, considering a finite but large $ J $ effect, one can further estimate the $ 1/(\mathfrak{p}^+)^2 $ correction which yields, $ \tilde{\gamma}_{c}= 2J -\frac{4\mathfrak{n}_{\nu}\omega}{\mathfrak{p}^+}+\frac{4\mathfrak{n}_{\mathfrak{K}}\omega^2}{(\mathfrak{p}^+)^2} +\cdots$. Clearly, $ 1/(\mathfrak{p}^+)^2 $ correction comes with the occupation number ($ \mathfrak{n}_{\mathfrak{K}} $) corresponding to massive modes. A similar expansion in (\ref{eqn42}) reveals, $ \Delta^{(c)}_S= \frac{\mathfrak{p}^+ J}{2 \omega}-\mathfrak{n}_{\nu}+ \frac{\mathfrak{n}_{\mathfrak{K}}\omega}{\mathfrak{p}^+}+\cdots$. 

These results suggest that both the massless as well as the massive modes start contributing to the spectrum as $ 1/J $ correction. Clearly, in the strict $ J=\infty $ limit all these corrections go away. We identify this as the point particle limit of the semiclassical string where the string fluctuations are all suppressed. 
\subsection{Beyond pp wave dynamics}
We now go beyond the standard pp wave limit \cite{Callan:2003xr}-\cite{Callan:2004uv} and consider curvature corrections to the metric upto an order $ \sim \mathcal{O}( \ell^{-2}) $ namely,
\begin{eqnarray}
ds^2 = ds^2_{pp}+\ell^{-1}ds^{2(1)}+\ell^{-2}ds^{2(2)}+\mathcal{O}(\ell^{-3})
\end{eqnarray}
where, the first order correction to the metric could be formally expressed as,
\begin{eqnarray}
ds^{2(1)}=\frac{2 \tilde{\gamma}}{\omega}d\mathfrak{x}^{+}d\eta
\end{eqnarray}

On the other hand, we enumerate the second order corrections to the metric as,
\begin{eqnarray}
ds^{2(2)}=\left( 2 \beta ^2 -\frac{2 \tilde{\gamma}}{\omega^{2}}\right) d\mathfrak{x}^{+} d\mathfrak{x}^{-} -\frac{ \omega ^2}{3}\left( \varrho ^4-\beta^4 +\frac{6 \tilde{\gamma}( \beta^2 - \varrho^2)}{\omega^2}\right)  (d\mathfrak{x}^{+})^{2} \nonumber\\+\frac{2 \tilde{\gamma}\beta^2}{\omega}d\mathfrak{x}^{+}d\Phi +(\text{d$\varrho $}^2 +\text{d$\eta $}^2 ) \varrho ^2
-\frac{\beta ^4 \text{d$\Phi $}^2}{3}+\frac{(d\mathfrak{x}^{-})^{2}}{\omega ^2}.
\end{eqnarray}

Finally, the curvature corrections to the \emph{normalized} NS-NS flux could be expressed as,
\begin{eqnarray}
\mathcal{B}_{2}=\mathcal{B}_{2pp}+\ell^{-1}\mathcal{B}_{2}^{(1)} +\ell^{-2}\mathcal{B}_{2}^{(2)}+\mathcal{O}(\ell^{-3})
\label{e39}
\end{eqnarray}
where the individual terms in the expansion (\ref{e39}) are noted to be,
\begin{eqnarray}
\mathcal{B}_{2}^{(1)}  &=&\frac{1}{\omega}d\mathfrak{x}^{+}\wedge d\eta \\
\mathcal{B}_{2}^{(2)}&=& -\frac{1}{2 \omega^3}d\mathfrak{x}^{-}\wedge d \Phi - \frac{\beta^2}{\omega}d\mathfrak{x}^{+}\wedge d \Phi.
\end{eqnarray}

As a result of curvature corrections, the corresponding Lagrangian density (\ref{e14}) 
\begin{eqnarray}
\mathcal{L}= 2\partial_{\tau}\mathfrak{x}^{-}+(\beta^2 \omega^2 - 2 \tilde{\gamma})-((\partial_{\tau}\eta)^{2}-(\partial_{\sigma}\eta)^2)-((\partial_{\tau}\beta)^{2}-(\partial_{\sigma}\beta)^2)\nonumber\\
-\beta^2 ((\partial_{\tau}\Phi)^{2}-(\partial_{\sigma}\Phi)^2)-\Delta \mathcal{L}
\end{eqnarray}
changes by an amount,
\begin{eqnarray}
\Delta \mathcal{L}=\frac{2 \tilde{\gamma}}{\omega \ell}\partial_{\tau}\eta +\frac{2}{\ell^2} \left(  \beta ^2 -\frac{ \tilde{\gamma}}{\omega^{2}}\right)\partial_{\tau}\mathfrak{x}^{-}+\frac{2 \tilde{\gamma}\beta^2}{\omega \ell^2}\partial_{\tau}\Phi +\frac{\varrho^{2}}{\ell^2}((\partial_{\tau}\varrho)^2 +(\partial_{\tau}\eta)^2)\nonumber\\
-\frac{\beta^{4}}{3 \ell^2}(\partial_{\tau}\Phi)^2 +\frac{1}{\omega^2 \ell^2}(\partial_{\tau}\mathfrak{x}^{-})^2-\frac{\varrho^{2}}{\ell^2}((\partial_{\sigma}\varrho)^2 +(\partial_{\sigma}\eta)^2)+\frac{\beta^{4}}{3 \ell^2}(\partial_{\sigma}\Phi)^2\nonumber\\
-\frac{1}{\omega^{2}\ell^2}(\partial_{\sigma}\mathfrak{x}^{-})^2-\frac{ \omega ^2}{3 \ell^2}\left( \varrho ^4-\beta^4 +\frac{6 \tilde{\gamma}( \beta^2 - \varrho^2)}{\omega^2}\right)-\frac{\beta^2}{\omega \ell^2}\partial_{\sigma}\Phi 
\label{e42}
\end{eqnarray}
where we drop total derivative terms those coming from the NS sector in the above expansion (\ref{e42}).

The corresponding canonically conjugate momenta are given by,
\begin{eqnarray}
\mathfrak{p}_{-}&=&2 -\frac{2}{\ell^2}\left(  \beta ^2 -\frac{ \tilde{\gamma}}{\omega^{2}}\right)-\frac{2}{\omega^2 \ell^2}\partial_{\tau}\mathfrak{x}^{-}\\
\label{e45}
\mathfrak{p}_{\eta}&=&-2 \partial_{\tau}\eta - \frac{2 \tilde{\gamma}}{\omega \ell}-\frac{2 \varrho^2}{\ell^2}\partial_{\tau}\eta\\
\mathfrak{p}_{\varrho}&=&-\frac{2 \varrho^2}{\ell^2}\partial_{\tau}\varrho\\
\mathfrak{p}_{\beta}&=&-2 \partial_{\tau}\beta \\
\mathfrak{p}_{\Phi}&=&- 2\beta^2 \partial_{\tau}\Phi - \frac{2 \tilde{\gamma}\beta^2}{\omega \ell^2}+\frac{2\beta^4}{3\ell^2}\partial_{\tau}\Phi.
\label{e48}
\end{eqnarray}

Notice that, $ \eta $ is an isometry of the target space hence $ \mathfrak{p}_{\eta} $ is conserved. This implies that $ \varrho = \varrho (\sigma) $ and therefore $ \mathfrak{p}_{\varrho}=0 $. On the other hand, inverting (\ref{e45}) and (\ref{e48}) we find,
\begin{eqnarray}
\partial_{\tau}\eta &=&-\frac{\tilde{\gamma}}{\omega \ell}-\frac{1}{2}\left(1-\frac{\varrho^2}{\ell^2} \right)\mathfrak{p}_{\eta} +\mathcal{O}(\ell^{-3})\\
\partial_{\tau}\Phi &=&-\frac{\mathfrak{p}_{\Phi}}{2 \beta^2}\left( 1+\frac{\beta^2}{3 \ell^2}\right)-\frac{\tilde{\gamma}}{\omega \ell^2}+\mathcal{O}(\ell^{-3}). 
\end{eqnarray}

Taking all these as inputs, the corresponding shift in the Hamiltonian spectrum turns out to be,
\begin{eqnarray}
\Delta \mathfrak{H}=\frac{1}{4 \pi \omega} \int_{0}^{2\pi} d \sigma \Delta \mathcal{H}
\end{eqnarray}
where,
\begin{eqnarray}
\Delta \mathcal{H}=\frac{\mathfrak{p}^2_{\Phi}}{2 \beta^2}\left(1+\frac{\beta^2}{2\ell^2} \right) +\left( 1-\frac{\varrho^2}{3\ell^2}\right) \mathfrak{p}^2_{\eta}+\frac{2\tilde{\gamma}\mathfrak{p}_{\eta}}{\omega \ell}+\frac{2\tilde{\gamma}\mathfrak{p}_{\Phi}}{\omega \ell^2}+\frac{2 \tilde{\gamma}^2}{\omega^2 \ell^2}\nonumber\\
+\frac{1}{\omega^2 \ell^2}\left( \frac{1}{8}\sum_{i=\eta , \Phi , \beta}\mathfrak{p}^2_i+\frac{1}{2}\sum_{i=\eta , \Phi , \beta}(\partial_{\sigma}X_i)^2-\frac{1}{2}(\omega^2 \beta^2 -2 \tilde{\gamma}) \right)^2\nonumber\\
-\frac{\varrho^{2}}{\ell^2}((\partial_{\sigma}\varrho)^2 +(\partial_{\sigma}\eta)^2)+\frac{\beta^{4}}{3 \ell^2}(\partial_{\sigma}\Phi)^2\nonumber\\
-\frac{1}{\omega^{2}\ell^2}(\partial_{\sigma}\mathfrak{x}^{-})^2-\frac{ \omega ^2}{3 \ell^2}\left( \varrho ^4-\beta^4 +\frac{6 \tilde{\gamma}( \beta^2 - \varrho^2)}{\omega^2}\right)-\frac{\beta^2}{\omega \ell^2}\partial_{\sigma}\Phi.
\end{eqnarray}
\subsection{A note on conformal gauge}
So far our analysis was performed using a particular gauge namely the light cone gauge. However, in this part of our analysis, we choose to work with a different gauge fixing procedure known as the \emph{conformal} gauge.  

In conformal gauge, one fixes both the re-parametrization and the Weyl invariance of the sigma model by choosing the world-sheet metric in the 2D Minkowskian form,
\begin{eqnarray}
\gamma_{\alpha \beta}=diag (-1, ~1).
\end{eqnarray}

This leads to the pp wave Polyakov action as
\begin{eqnarray}
\mathcal{S}_{pp}=\frac{1}{4 \pi \alpha'}\int d\tau d\sigma ~ \mathfrak{L}_{pp}~,
\end{eqnarray}
where we identify the corresponding Lagrangian density as
\begin{eqnarray}
 \mathfrak{L}_{pp}=-2 (\partial_{\tau}\mathfrak{x}^{+}\partial_{\tau}\mathfrak{x}^{-}-\partial_{\sigma}\mathfrak{x}^{+}\partial_{\sigma}\mathfrak{x}^{-})-( \omega ^2 \mathcal{X}^2_a  - 2 \tilde{\gamma} )((\partial_{\tau}\mathfrak{x}^{+})^2 - (\partial_{\sigma}\mathfrak{x}^{+})^2)\nonumber\\
 +((\partial_{\tau}\mathcal{Z}_i)^2 -(\partial_{\sigma}\mathcal{Z}_i)^2).
\end{eqnarray}
\subsubsection{Dynamics}
The corresponding conjugate momenta are given by,
\begin{eqnarray}
\mathfrak{p}_{+}&=&-2 \partial_{\tau}\mathfrak{x}^{-}+( \omega ^2 \mathcal{X}^2_a  - 2 \tilde{\gamma} )\mathfrak{p}_{-}\\
\mathfrak{p}_{-}&=&-2  \partial_{\tau}\mathfrak{x}^{+}\\
\mathfrak{p}_{i}&=&2 \partial_{\tau}\mathcal{Z}_i.
\end{eqnarray}

As $ \mathfrak{p}_- $ is a constant of motion therefore we find,
\begin{eqnarray}
\partial^2_{\tau}\mathfrak{x}^{+}=0
\end{eqnarray}
whose most general solution can be expressed as,
\begin{eqnarray}
\mathfrak{x}^{+}= \left( \frac{\mathfrak{p}^+}{\omega}\right) \tau +\mathfrak{a}^+
\label{e61}
\end{eqnarray}
where we identify $ \mathfrak{p}^+ $ as the momentum of the string along the light cone direction $ \mathfrak{x}^+ $. 

We fix the above entities $ \frac{\mathfrak{p}^+}{\omega} $ and $ \mathfrak{a}^+(\sigma) $ such that one ends up in a solution of the form, $ \mathfrak{x}^{+}=\tau $ which is precisely the gauge fixing condition in the light cone gauge. In other words, we show the equivalence between two different gauge choices. Following the above set of arguments, one may therefore conclude that the Hamiltonian spectrum does not depend on the choice of gauge. In other words, it is gauge invariant.

Finally, we note down the corresponding pp wave Hamiltonian density,
\begin{eqnarray}
\mathcal{H}^{(cg)}_{pp}=-\frac{\mathfrak{p}_- \mathfrak{p}_+}{2}+\mathfrak{p}_- \mathcal{H}^{(lc)}_{pp}-2\partial_{\sigma}\mathfrak{x}^+ \partial_{\sigma}\mathfrak{x}^- -( \omega ^2 \mathcal{X}^2_a  - 2 \tilde{\gamma} )(\partial_{\sigma}\mathfrak{x}^+)^2.
\end{eqnarray}
\subsubsection{Gauge fixing}
In order to check the residual gauge freedom of the model, we adopt the following change in coordinates 
\begin{eqnarray}
\sigma^{\pm}=\tau \pm \sigma ,
\end{eqnarray}
which yields the pp wave Lagrangian density of the following form,
\begin{eqnarray}
\mathfrak{L}_{pp}=-4(\partial_+ \mathfrak{x}^+ \partial_- \mathfrak{x}^- +\partial_+\mathfrak{x}^-\partial_-\mathfrak{x}^+)-4( \omega ^2 \mathcal{X}^2_a  - 2 \tilde{\gamma} )\partial_+ \mathfrak{x}^+ \partial_- \mathfrak{x}^++4\partial_+ \mathcal{Z}_i \partial_-\mathcal{Z}_i.
\end{eqnarray}

Clearly, as a result of this, the Polyakov action
\begin{eqnarray}
\mathcal{S}_{pp}=\frac{1}{8 \pi \alpha'}\int d\sigma^+ d\sigma^- \mathfrak{L}_{pp} 
\end{eqnarray}
is invariant under the following rescaling,
\begin{eqnarray}
\sigma^{\pm}=\tilde{\sigma}^{\pm}(\sigma^{\pm}).
\end{eqnarray}

Therefore, we are left with two more gauge freedoms one of which could be fixed by setting,
\begin{eqnarray}
\partial_{\sigma}\mathfrak{x}^+ = 0
\end{eqnarray}
which sets, $ \mathfrak{a}^+ =0 $ in (\ref{e61}). 

Using, the remaining gauge freedom, one can further fix the dimensionless ratio, 
\begin{eqnarray}
 \frac{\mathfrak{p}^+}{\omega}=1
\end{eqnarray}
 which naturally fixes the pp wave Hamiltonian in the canonical gauge as, $ \mathcal{H}^{(cg)}_{pp}=-\mathfrak{p}_+ \equiv \mathcal{H}^{(lc)}_{pp}$ and therefore the spectrum turns out to be identical in both descriptions. 
\section{Summary and final remarks}
We conclude our analysis with a brief summary of the results. In the present paper, using light cone gauge, we estimate pp wave stringy spectrum over massless $ BTZ\times S^3 $ in the presence of NS-NS fluxes. It turns out that the pp wave spectrum receives a constant energy shift ($ \Delta_S $) due to TsT transformations acting along $ U(1) $ isometries of the target space manifold. More precisely, this shift appears to be proportional to the string momentum ($ \mathfrak{p}^+ $) along light cone direction $ \mathfrak{x}^+ $. Imposing the energy positivity constraint on the spectrum ($ \Delta >0 $), we further propose an upper bound ($ \Delta^{(c)}_S $) on the energy shift in the limit of large $ R $- charge.  We also estimate $ \ell^{-2} $ corrections to the spectrum as we go beyond the strict pp wave limit. Finally, we compute the pp wave Hamiltonian using conformal gauge and show the equivalence between two approaches. Before we conclude, it is worthwhile to mention that a parallel derivation of similar results from CFT$ _2 $ data would certainly be an exciting direction to be explored in the near future. \\\\
{\bf {Acknowledgements :}}
 The author is indebted to the authorities of IIT Roorkee for their unconditional support towards researches in basic sciences. The author would also like to acknowledge The Royal Society, UK for financial assistance, and acknowledges the Grant (No. SRG/2020/000088) received from The Science and Engineering Research Board (SERB), India.

\end{document}